\PassOptionsToPackage{dvipsnames,table,xcdraw}{xcolor}
\documentclass[sigconf]{acmart}

\usepackage{graphicx}
\usepackage{multirow}
\usepackage{subcaption}
\usepackage{makecell}
\usepackage{pbox}
\usepackage{amsmath}
\begin{document}


%\title{Assisting Ad Creative Design via Visual-Linguistic Representations: A Keyphrase Recommendation Approach}

\title{Powering COVID-19 community Q\&A with Curated Side Information}

\author{Manisha Verma, Kapil Thadani, and Shaunak Mishra}
\affiliation{
 \institution{Yahoo! Research, VerizonMedia, New York}}
  \email{manishav,thadani,shaunakm@verizonmedia.com}

\begin{abstract}
Community question answering and discussion platforms such as Reddit, Yahoo! answers or Quora provide users the flexibility of asking open ended questions to a large audience, and replies to such questions maybe useful both to the user and the community on certain topics such as health, sports or finance. Given the recent events around COVID-19, some of these platforms have attracted 2000+ questions from users about several aspects associated with the disease. Given the impact of this disease on general public, in this work we investigate ways to improve the ranking of user generated answers on COVID-19. We specifically explore the utility of external technical sources of side information (such as CDC guidelines or WHO FAQs) in improving answer ranking on such platforms. We found that ranking user answers based on question-answer similarity is not sufficient, and existing models cannot effectively exploit external (side) information. In this work, we demonstrate the effectiveness of different attention based neural models that can directly exploit side information available in technical documents or verified forums (e.g., research publications on COVID-19 or WHO website). Augmented with a temperature mechanism, the attention based neural models can selectively determine the relevance of side information for a given user question, while ranking answers.
%and our experiments with fine tuned models such as BERT showed that large 
\end{abstract}

\copyrightyear{2019} 
\acmYear{2019} 
%\acmConference[WWW]{World Wide Web}{2020}{USA}

\maketitle

\begin{comment}
We study the possibility of using open research datasets for COVID-19 question answering (QA) in the context of Yahoo Answers. In particular, given a COVID-19 related question, we focus on predicting the top answer (highest upvotes) from the observed list of answers. We demonstrate that side information available in technical documents (\emph{e.g.}, research publications on COVID-19) can be helpful towards the above objective when carefully used with an attention (gating) mechanism. Our experiments show how certain categories of questions (all related to COVID-19) can significantly benefit from such side information.
\end{comment}

\section{Introduction} \label{sec:introduction}

Question answering systems are key to finding relevant and timely information about several issues. Community question answering (cQ\&A) platforms such as Reddit, Yahoo! answers or Quora have been used to ask questions about wide ranging topics. Most of these platforms let users ask, answer, vote or comment on questions present on the platform. However, question answering platforms are useful not only for getting public opinions or votes about areas such as entertainment or sports but can also serve as information hot-spots for more sensitive topics such as  health, injuries or legal topics. Thus, it is imperative that when the user visits  
\emph{sensitive} topics content, answer ranking also takes into account curated side information from reliable (external) sources. Most prior work on cQ\&A has focused on incorporating question-answer similarity \cite{LTR_2008,semantic_sim2015}, user reputation \cite{yang2016beyond,hong2009classification,user_interaction_2013}, integration of multi-modal content \cite{medical_qa2019}, community interaction features \cite{comm_interaction2018} associated with answers or just the question answering network \cite{comm_net2019} on the platform. However, there is very limited work on incorporating \emph{curated content} from external sources. Existing work only exploits knowledge bases \cite{medical_qa2019} that consist of different entities and relationships between these entities to score answers. However, there are some limitations of knowledge bases that would make it difficult to use them for community Q\&A for rapidly evolving topics such as disease outbreaks (e.g. ebola, COVID-19), wild-fires or earthquakes.  
\begin{figure}[t]
  \includegraphics[width=0.45\textwidth]{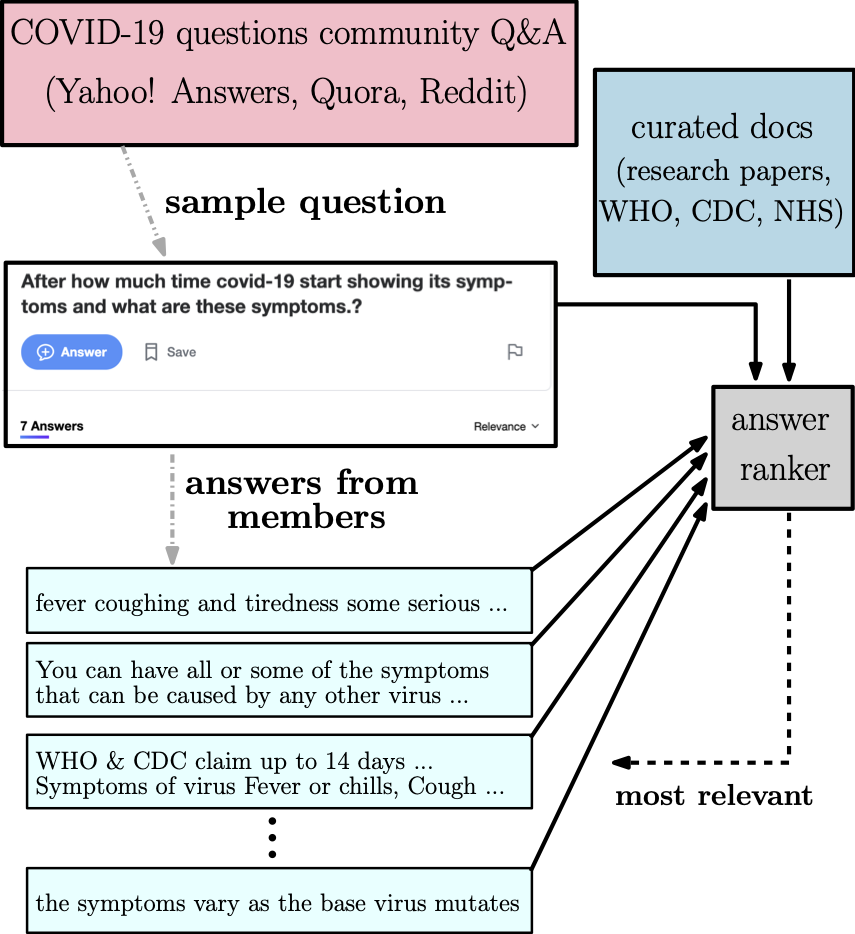}
  \caption{Illustrative example of COVID-19 community answer ranking powered by side information in the form of research papers, and information from verified sources (such as CDC, WHO, and NHS).}
  \label{fig:cques}
\end{figure}
Firstly, knowledge bases contain information about established entities, and do not rapidly evolve to incorporate new information which makes them unreliable for novel disease outbreaks such as COVID-19 where information rapidly changes and its verification is time sensitive. Secondly, it may be hard to determine what even \emph{constitutes an entity} as new information arrives about the topic. To overcome these limitation in this work, we posit that external \emph{curated} free-text or semi-structured informational sources can also be used effectively for cQ\&A tasks. 

In this work, we demonstrate that free text or semi structured external information sources such as CDC\footnote{https://www.cdc.gov/}, WHO\footnote{https://www.who.int/} or NHS\footnote{https://www.nhs.uk/} can be very useful for ranking answers on community Q\&A platforms since they contain frequently updated information about several topics such as ongoing disease outbreaks, vaccines or resources about other topics such as surgeries, birth control or historical numerical data about diseases across the world.

We argue that for sensitive topics such as COVID-19, it is useful to use publicly available \emph{vetted} information for improving our ranking systems. In this work, we explore the utility of publicly available information for ranking answers for questions associated with COVID-19. We specifically focus on ranking answers for questions in two publicly available \emph{primary} Q\&A datasets: a) Yahoo! Answers\footnote{https://answers.yahoo.com/} and b) recently released annotated Q\&A dataset \cite{emnlp_data2020} in presence of two \emph{external} semi-structured curated sources: a) TREC-COVID \cite{voorhees2020treccovid} and b) WHO questions and answers \footnote{https://www.who.int/emergencies/diseases/novel-coronavirus-2019/question-and-answers-hub} on COVID-19. We explore the utility of deep learning models with attention in this work to improve upon existing state-of-the-art systems. 

More specifically, we propose a temperature regulated attention mechanism to rank answers in presence of external (side) information. Our experiments on 10K+ questions from both \emph{source} datasets on COVID-19 show that our models can improve ranking quality by a significant margin over question-answer matching baselines in presence of external information. Figure \ref{fig:cques} demonstrates the overall design of our system. We specifically use attention based neural architecture with temperature to automatically determine which components in the external information are useful for ranking user answers with respect to a question. Ranking performance, when evaluated with three metrics shows that precision and recall for correct answer retrieval improves by $\sim$17\% and $\sim$9\% for both source datasets respectively over several other cQ\&A models. 

%Category based analysis of performance further indicates superiority of the proposed model over text (or embedding) based matching of question-answer.  

%has received over 2000+ questions related to various aspects of the virus since the beginning of January 2020. A small sample of three questions is shown in Figure \ref{fig:cques} where users are asking about origins and self diagnosis of the virus. 

\section{Related work} \label{sec:related}
Community Question and Answering (cQ\&A) systems is a well researched sub-field both in information retrieval and NLP communities. Several systems 
have been proposed to rank user submitted answers to questions on community platforms such as Yahoo! answers, Reddit and Quora. 

Ranking user submitted answers on community question-answering platforms has been addressed with several approaches. Primary method is to determine the relevance of the answer given an input question. Text based matching is one of the most common approaches to rank answers. Researchers have used several methods to compute \emph{similarity} between a question and user generated answers to determine relevance. For instance, feature based question-answer matching is used in \cite{LTR_2008} with 17 features extracted from unigrams, bigrams and web correlation features using unstructured user search logs to rank answers. It is worth noting that user features and community features when incorporated may still yield further improvements in the performance of these models but this is not the focus of our work. The authors in \cite{LTR_2008} used questions extracted from Yahoo! answers for their experiments. Researchers have used different approaches such representation learning, for instance, in \cite{LSTMans2017IWCS,cohen_2016} authors use LSTM to represent questions and answers respectively. Convolutional networks have also been used in \cite{yang2016beyond,zhou2015answer} to rank answers. Other approaches such as doc2vec \cite{nie2017data}, tree-kernels \cite{severyn2012structural}, adversarial learning \cite{yang2019adversarial}, attention \cite{attention_question2018,medical_qa2019,LSTMans2017IWCS, attentive2017AAAI} or deep belief networks \cite{wang2010modeling} have been used to score question and answer pairs. There have also been studies exploring community, user interaction or question based features \cite{yang2016beyond, hong2009classification, user_interaction_2013,comm_interaction2018} to rank answers. While these approaches are relevant, it is not always evident how one can incorporate external information when it is either in free-text or semi-structured format into these systems. We explore some question-answer based matching approaches as baselines in this work and show that for rapidly evolving topics such as COVID-19, inclusion of external \emph{curated} information can boost model performance. 

The line of work most closely related to ours is incorporation of knowledge bases in Q\&A systems. Existing work \cite{rankQA_2019,kb_rank2018Sig,medical_qa2019}, however, approaches different tasks. For instance, authors in \cite{kb_rank2018Sig,rankQA_2019} focus on finding factual answers to questions using a knowledge base. This does not extend easily to cQ\&A where neither the questions nor the answers may request or refer to any facts. Most recent work is \cite{medical_qa2019} on incorporating medical KB for ranking answers on medical Q\&A platforms. They propose to learn path based representation of \emph{entities (from KB)} present in question and answers posted by users. This approach relies on reliable detection of entities first, which may be absent for emerging topics such as COVID-19 pandemic. Another limitation of this work is that external knowledge may not always be present in a \emph{structured} format. For example, CDC guidelines are usually simple question-answer pairs posted on the website. This makes it difficult to apply their approach to our problem. The proposed approach in this work incorporates semi-structured information directly with help of temperature regulated attention. 

Finally, with the rise of COVID-19, researchers across disciplines are actively publishing information and datasets to share understanding of the virus and its impact on people. Researchers routinely organize dedicated challenges such as SemEval \cite{nakov2019semeval} with tasks such as ranking answers on QA forums. One such initiative is TREC-COVID track \cite{voorhees2020treccovid} which released queries, documents and manual relevance judgements to power search for COVID related information. Authors in \cite{cairecovid} also released COVID-19 related QA dataset with 100+ questions and answers pairs extracted from TREC COVID \footnote{https://ir.nist.gov/covidSubmit/data.html} initiative. These questions/answer pairs are not user generated content, hence, do not reflect real user questions. We also rely on recently released Q\&A dataset from \cite{emnlp_data2020} for our task. We also compile a dataset of 2000+ COVID-19 questions with 10K+ answers all submitted by users on Yahoo! answers for this work. 

\begin{comment}
\subsection{QA on unstructured knowledge bases}
BioBERT \cite{BioBERT}
\cite{,}
Knowledge search \cite{sledge_scibert}.
\end{comment}

% https://www.aaai.org/ocs/index.php/AAAI/AAAI18/paper/download/17226/16146

% https://arxiv.org/pdf/1905.02019.pdf
% student report (perhaps not relevant) https://web.stanford.edu/class/archive/cs/cs224n/cs224n.1174/reports/2761224.pdf
% https://dl.acm.org/doi/10.1145/3077136.3080699 [sigir short position]
% https://dl.acm.org/doi/10.1145/3209978.3210081 [sigir short KB]
% https://dl.acm.org/doi/pdf/10.1145/3184558.3191830 [www short]
% https://dl.acm.org/doi/pdf/10.1145/3308558.3313518 [www short]
% https://www.aclweb.org/anthology/P16-1122.pdf
% https://www.aclweb.org/anthology/D18-1240.pdf 
% https://www.aclweb.org/anthology/C18-1279.pdf (CICLING)
% https://www.aclweb.org/anthology/C18-1215.pdf (CICLING)
% https://arxiv.org/pdf/1905.10720.pdf [arxiv long]

%\input{./sections/data.tex}
\section{Method} \label{sec:method}

\subsection{Problem formulation}
In this work, we focus on ranking answers for $n$ questions $q_1, \ldots, q_n$ related to an emerging topics such as COVID. Each $q_i$ is associated with a set of two or more answers
$A_i = \{a_{ij}: j \ge 2\}$ and corresponding labels 
$Y_i = \{y_{ij}: j \ge 2\}$ representing answer relevance. We use binary indicator for relevance where relevant judgments (e.g., \textit{favorite}, \textit{upvoted}) are provided by the user, i.e., $y_{ij} \in \{0, 1\}$ respectively. 

\begin{comment}
Labels may
be defined as either:
\begin{itemize}
    \item Binary relevant judgments (e.g., \textit{favorite}, \textit{selected}) provided by the questioner, i.e., $y_{ij} \in \{0, 1\}$
    \item Integers representing aggregated votes provided by the community, i.e., $y_{ij} \in \mathbb{Z}$.
\end{itemize}
\end{comment}
We attempt to model the relevance of each answer $a_{ij}$ to its corresponding question using an external source which may contain free text or semi-structured information. For example, the \emph{external} source could consist of information-seeking queries or questions $eq_1, \ldots, eq_m$ related to a topic, with each $eq_k$ linked to a set of relevant scientific articles or answers $ED_k$, where each answer/document $ed_1, \ldots, ed_p$ may be judged for relevance by human judges \cite{voorhees2020treccovid} or some experts. 

We hypothesize that this semi-structured or free-text information may be valuable in identifying user answer quality for certain kinds of questions, although not all. We investigate this with our model to recover the true labels $y_{ij}$ for each user answer $a_{ij} \in A_i$ given its question $q_i$, category information, and information from the \emph{external} source $\langle eq_k, ED_k \rangle_{k=1}^m$.

\subsection{Proposed Model}
\begin{figure}[t]
  \includegraphics[width=0.50\textwidth]{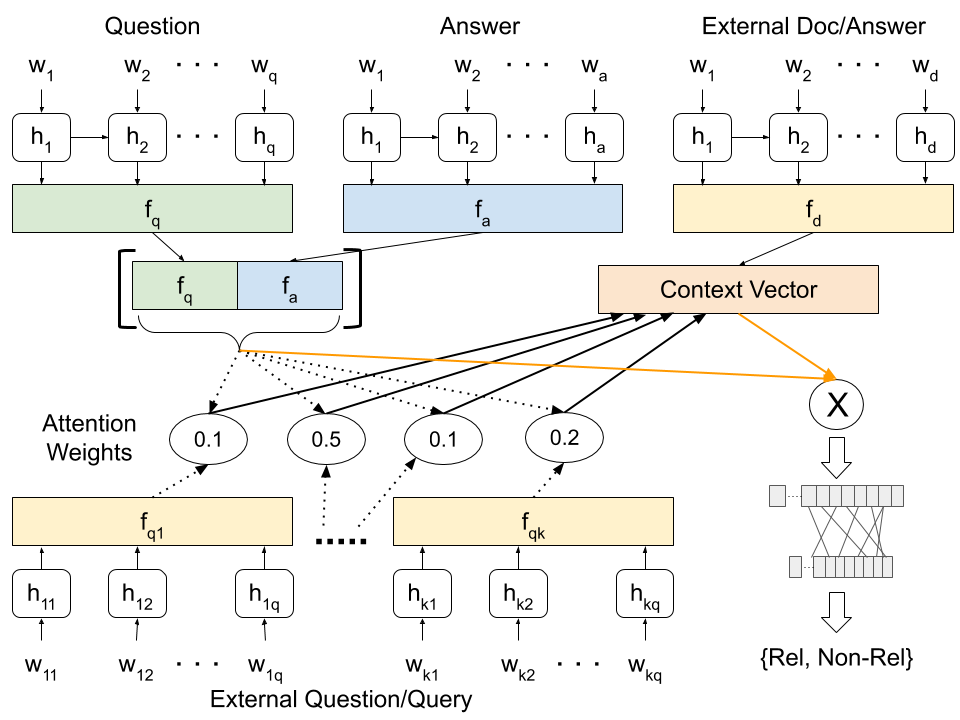}
  \caption{External source augmentation model}
  \label{fig:trec_doc}
\end{figure}

In this work, we explore token-level matching mechanism to determine the
relevance of information in the external source that may inform the label prediction task. Our model (\emph{$\tau$-att}) aims to match a given user question with all the submitted answers in the presence of external information about the same domain. 
First, the question $q_i$, an answer $a_{ij}$ and additional metadata can be
encoded into a $d$-dimensional vector $x_i$ using a text encoder
$f_\text{input}$. We use LSTM based encoder for both question and answer in the \emph{primary} source which can handle input sequences of variable length. 
\paragraph{Question Encoding:} Each word $w_i^q$ in a question is represented as a $K$ dimensional vector with pre-trained word embeddings. LSTM takes each token embedding as input and updates hidden state $h_i^q$ based on previous state $h_{i-1}^q$. Finally, the hidden state is input to a feed forward layer with smaller dimension $F<K$ to compress question encoding as follows:
\begin{equation}
    h_i^q = LSTM(h_{i-1}^q , w_i^q),  \: 
    f_i^q = RELU(h_i^q W_q + b_q)
\end{equation}

\paragraph{Answer Encoding:} Each word $w_j^a$ in the answer is also represented as a $K$ dimensional vector with pre-trained word embeddings. LSTM takes each token embedding as input and updates hidden state $h_j^a$. We also reduce the dimension of answer encoding with a feed forward layer with dimension $F<K$ as follows:
\begin{equation}
    h_j^a = LSTM(h_{j-1}^a , w_j^a),  \: 
    f_j^a = RELU(h_j^a W_a + b_{a})
\end{equation}
We concatenate the question and answer representations for further processing. 
\begin{equation}
    f_{ij} = [f_i^q,  f_j^a]
\end{equation}

\paragraph{External source encoding:} External sources of information can vary from task-to-task. We encode each segment of data individually. For instance, if there are two segments in the source (e.g. question/answer or query/document), our system encodes both segments individually. We use the same encoding architecture used for primary source question/answer encoding above. Encoding example for two segment \emph{external} source is given below.
\begin{equation}
\begin{split}
    h_t^{eq} = LSTM(h_{t-1}^{eq} , w_t^{eq}),  \: 
    f_t^{eq} = RELU(h_t^{eq} W_{eq} + b_{eq}) \\
    h_t^{ed} = LSTM(h_{t-1}^{ed} , w_t^{ed}),  \: 
    f_t^{ed} = RELU(h_t^{ed} W_{ed} + b_{ed}) \\
\end{split}
\end{equation}

%If no side information was available, this vector would normally be supplied
%directly to a label prediction network $g$.
\begin{table*}[th]
\begin{tabular}{|l||l|l|l|}\hline
 Source & Question & Rel answer & Non-rel answer   \\ \hline
\hline
Yahoo! Ans & \pbox{3cm}{I am really scared to go places for St. Patrick’s day because of the coronavirus. what do I do? }& \pbox{7cm}{Unfortunately, there’s not enough people that care and will still go out and party despite the coronavirus epidemic. I’m proud of you in that you’re taking extra precautions ... Good for you!} & \pbox{5cm}{Stop being scared of viruses. What's the problem?} \\ \hline
Infobot & \pbox{3cm}{Can corona live on cardboard? } & \pbox{7cm}{A recent study shows that the virus can live in the air ... On cardboard, it can live up to 24 hours (1 day)} & \pbox{5cm}{The risk is quite low for one to become infected with COVID19 through mail/packages - especially because...(over a period of a few days/weeks).}  \\ \hline
\end{tabular}
\caption{Sample rel/non-rel answers from both sources}
\label{table:sample data}
\end{table*}
We incorporate external source encoding with a temperature $(\tau)$ based variant of scaled dot-product attention, which provides a straightforward conditioning approach over a set of query-document pairs. Question encoding vector $f_{ij}$ serves as a query over keys $f_t^{eq}$. If two segments are present in the external source such as query/document, the model uses the attention weights over first segment (e.g. query) to determine the importance of the second segment (e.g. document) respectively. It is easy to extend this framework to \emph{external} sources with multiple segments. The two segment attention is described below. 
\begin{equation}
    z_{it} = \frac{f_{ij}^\top f_t^{eq}}{\sqrt{d}} \\
    \alpha_{it} = \frac{e^{z_{it}/\tau}}{\sum_l e^{z_l/\tau}} \\
    s_{itd}' = \sum_d \alpha_{it} f_t^{ed}
\end{equation}
To summarize, temperature $(\tau)$ based attention helps determine the relevance of each $f_t^{ed}$ corresponding $f_t^{eq}$ with respect to the question encoding. Temperature $(\tau)$ parameter helps us control the uniformity of attention weights $\alpha_{it}$. Finally, labels are predicted using a multi-layer perceptron over the input vector $f_{ij}$ and the learned weighted average of side information $s_{itd}'$. We use binary cross entropy loss to train the proposed model. 
\begin{equation}
    \hat{y}_{ij} = F_\text{output}([f_{ij}; s_{itd}'])
\end{equation}
where $F_\text{output}$ uses sigmoid activation function. Since community questions may often be entirely unrelated to external sources, a key aspect of this approach is determining \textit{whether}
the \emph{external source} is useful, not merely attending to its entries that are most relevant. Temperature based attention mechanism is useful in controlling which external source entries are useful for user  questions. It is worth noting that one will have to experiment and tune the value of temperature $\tau$ such that ranking performance improves. 

%One simple extension of our model would be to introduce a dummy entry $\langle r_{m+1}, \S_{m+1} \rangle$ to the side information where the corresponding $s_{m+1}$ is a zero vector that produces no activations within $f_\text{output}$. This, however, is left for future work. 
\begin{comment} 
each expanded query $r_k$ and
corresponding collection of scientific abstracts $S_k$ from the TREC dataset are
also encoded into vectors using pretrained encoders.
\begin{align*}
    p_k &= f_\text{query}(r_k) \\
    s_k &= f_\text{docs}(S_k)
\end{align*}
\kapil{note that an average may be too coarse here.}
\end{comment}
\begin{comment}
\subsection{Gating side-information}
We consider two strategies for achieving this:
\begin{enumerate}
    \item Learning a softmax temperature parameter $t$ as a function of $x_{ij}$
        in the attention mechanism, which can prevent the softmax operation from
        concentrating attention weights for non-scientific questions.
        \begin{align*}
            t &= f_\text{temp}(x_{ij})
        \end{align*}
    \item Adding a dummy entry $\langle r_{m+1}, \S_{m+1} \rangle$ to the side
        information where the corresponding $s_{m+1}$ is a zero vector that
        produces no activations within $f_\text{output}$.
\end{enumerate}
\end{comment}

\section{Experimental Setup}
Given the model architecture, in this section, we provide a detailed overview of different datasets, metrics and baselines used in our experiments. 
\begin{figure}[t]
    \centering
    \begin{subfigure}[b]{0.49\columnwidth}
        \includegraphics[width=\textwidth]{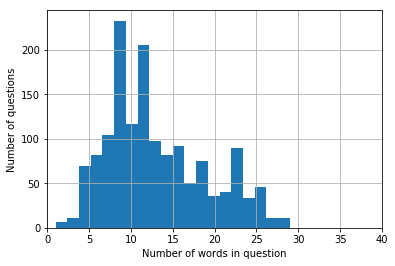}
    \caption{Yahoo! ques length}
    \label{fig:yahoo_qlen}
    \end{subfigure}
    \begin{subfigure}[b]{0.50\columnwidth}
        \includegraphics[width=\textwidth]{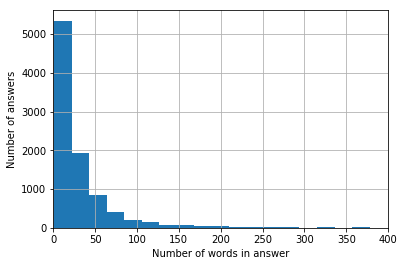}
        \caption{Yahoo! ans length}
        \label{fig:yahoo_alen}
    \end{subfigure}
    \begin{subfigure}[b]{0.49\columnwidth}
        \includegraphics[width=\textwidth]{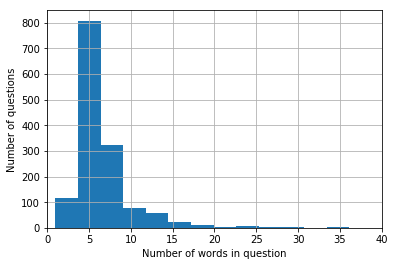}
        \caption{Infobot ques length}
        \label{fig:emnlp_qlen}
    \end{subfigure}
     \begin{subfigure}[b]{0.49\columnwidth}
        \includegraphics[width=\textwidth]{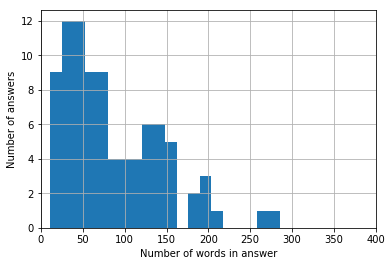}
        \caption{Infobot ans length}
        \label{fig:emnlp_alen}
    \end{subfigure}
    \caption{Token distribution in different sources}
\end{figure}
\begin{comment}
\begin{figure}[t]
  \includegraphics[width=0.50\textwidth]{images/trec_query1.png}
  \caption{Sample trec query, narrative and description}
  \label{fig:trec_query}
\end{figure}
\end{comment}
\subsection{Data}
\label{subsec:data}
We compiled two question answering datasets. The first was collected from Yahoo! answers and the second was recently released in \cite{emnlp_data2020} where both datasets have questions raised by real users. In this work we focus specifically on questions associated with COVID-19. 
Different statistics about the train and test split of both q\&a datasets are given in Table \ref{table:train_test_stat} respectively. A pair of relevant and non-relevant answers for a question in both datasets is also shown in Table \ref{table:sample data} for reference. More details about them is given below. 
\begin{table}[t]
\begin{tabular}{|l|l|l|}\hline
Stat & Yahoo! Ans & Infobot \\ \hline 
Train Q-A & 9341& 6354\\ \hline
Train ans/q & 6.25$\pm$2.9 & 4.40$\pm$0.77 \\ \hline
Train \#qwords &12.71$\pm$5.8 &6.55$\pm$3.93 \\ \hline
Train \#awords &36.31$\pm$93.59 & 92.17$\pm$59.27 \\ \hline
Test Q-A & 2232&1592 \\ \hline
Test ans/q & 5.96$\pm$2.87 & 4.41$\pm$0.76 \\ \hline
Test \#qwords &13.07$\pm$5.89 &6.21$\pm$2.94 \\ \hline
Test \#awords &35.64$\pm$80.31 &92.39$\pm$59.47 \\ \hline
\end{tabular}
\caption{Train and test data from primary sources}
\label{table:train_test_stat}
\end{table}
\paragraph{Yahoo! Dataset}: We crawled COVID-19 related questions from Yahoo! answers \footnote{https://answers.search.yahoo.com/search?p=coronavirus} using several keywords such as `coronavirus', `covid-19', `covid', `sars-cov2' and `corona virus' between the period of Jan 2020 to July 2020 to ensure we gather all possible questions for our experiments. We keep only those questions have two or more answers. In total, we obtained 1880 questions with 11500 answers. We used favorite answers as positive labels (similar to previous work \cite{LTR_2008}), assuming that users, over time rate answers (with upvotes/downvotes) that are most relevant to the submitted question. We normalized the question and answer text by removing a small list of stop words, numbers, links or any symbols. Figure \ref{fig:yahoo_qlen} and \ref{fig:yahoo_alen} show the distribution of question and answer lengths respectively. Questions contain $12.7\pm5.8$ \emph{(qwords)} words and answers consist of $36.3\pm93.5$ (mean$\pm$std) words \emph{(awords)} respectively which indicates that user submitted answers can vary widely on Yahoo! answers. On average, a question has about 6 answers \emph{(ans/q)} in Yahoo! ans dataset. We spilt the data into three sets: train (64\%, 1196 questions, 7435 answers), validation (16\%, 298 questions, 1858 answers) and test (20\%, 374 questions, 2310 answers) set where questions for each set were uniformly sampled. 
%To construct the ground truth for our experiments, we label the favorite answer i.e. the top rated answer as the correct answer. It is worth noting that factors beyond the quality of the answer, such as upvotes, downvotes or user comments may influence the position of an answer with respect to a question on Yahoo! answer forums. In the absence of manually labeled data about factual correctness and relevance of the answers to COVID-19 questions, we rely on the favorite answer as our ground truth. 
\paragraph{Infobot Dataset \cite{emnlp_data2020}}: Researchers at JHU \cite{emnlp_data2020} have recently compiled a list of user submitted questions on different platforms and manually labeled 22K+ question-answer pairs. We cleaned this set by removing questions with less than two answers or no relevant answers. In total, our dataset contains 8000+ question answer pairs where each question may have \emph{multiple} relevant answers which is not the same as Yahoo! answers dataset. Figure \ref{fig:emnlp_qlen} and \ref{fig:emnlp_alen} show the distribution of question and answer lengths respectively. 
%We found that the answers in Infobot data are longer than those in Yahoo! answers data but question lengths are comparable across both datasets. 

%Query text can be expanded to include alternative formulations such as a question or a short narrative describing the informational goals of the searcher.

%In this work, we have two kinds of sources: curated question-answer and query-document list. For the latter, we also have relevance 

%Here, $f_\text{query}$ may be the same generic encoder as $f_\text{input}$ or a
%different encoder that produces $d$-dimensional embeddings. In contrast, $f_\text{docs}$ should be adapted to scientific literature, so we produce abstract embeddings using SPECTER \cite{cohan2020specter} and combine the embeddings of the most query-relevant documents.

%In this work, we use LSTM, followed by a feed-forward layer to encode yahoo questions, answers and trec queries, narratives and description. We use SPECTER embeddings of abstracts \footnote{semanticscholar.org/cord19/download}, followed by a feed-forward layer to represent documents. 
 
\subsubsection{External sources}
We use two external datasets to rank answers. Details of each dataset are given below:
\paragraph{TREC COVID \cite{voorhees2020treccovid}: } 
We use recently released TREC COVID-19 track data with 50 queries which also contain manually drafted query descriptions and narratives. Expert judges have labeled over 5000 scientific documents for these 50 queries from the CORD-19 dataset \footnote{https://www.semanticscholar.org/cord19}. These documents contain coronavirus related research. Given the documents are scientific literature, we initialize document embeddings using SPECTER \cite{cohan2020specter}.
\begin{comment}
\begin{figure}[t]
  \includegraphics[width=0.50\textwidth]{images/trec_doc.png}
  \caption{Sample document abstract labeled for trec}
  \label{fig:trec_doc}
\end{figure}
\end{comment}

\paragraph{WHO:} We use data released on question and answer hub of WHO website\footnote{https://www.who.int/emergencies/diseases/novel-coronavirus-2019/question-and-answers-hub} to create a list of question-answer pairs. There are 147 question and answer pairs in this dataset where questions contain 13.28$\pm$5.36 words and answers contain 133.2$\pm$100.9 words respectively. 

\subsection{Baselines}
\label{subsec:baselines}
We evaluated our model against embedding similarity baseline. We computed four baselines as follows:
\paragraph{\textbf{Random:}} An answer is chosen at random as relevant for a user question. This is expected to provide a lower bound on retrieval performance.  
\paragraph{\textbf{Linear Attention (\emph{att})}}: When $\tau=1.0$, our model defaults to simple linear attention over all the information present in the external sources. This gives an indication of how well the model performs when its forced to look at all the information in the external source. 
\paragraph{\textbf{Linear combination (\emph{$\lambda$-sim})}}: We linearly combine similarities between Yahoo! question-answer and Trec query-answer as shown below:
    \begin{equation}
            \emph{$\lambda$-sim} = \lambda \; cos(ya,yq) + (1-\lambda) \; \max_{tq} (cos(ya, tq)) 
    \end{equation}
    where $ya$, $yq$ and $tq$ are Yahoo! answer, question and concatenated trec query, narrative and description embeddings respectively. This is a more crude version of temperature attention where $\lambda$ controls the contribution of each component directly. We vary $\lambda$ to determine the optimal combination. Question-Answer similarity (\emph{qasim}) is similarity between question and answer embedding i.e. $\lambda=1$. Both question and answer embeddings are obtained by averaging over their individual token embeddings. 
\paragraph{\textbf{BERT Q\&A (\emph{bert})}}: Large scale pre-trained transformers \cite{devlin2019bert} are widely popular for NLP tasks. BERT like models have shown effectiveness on Q\&A datasets such as SQUAD \footnote{https://rajpurkar.github.io/SQuAD-explorer/}. We fine-tune BERT base model with two different answer lengths a) 128 \emph{(bert-sl128)} and b) 256 tokens \emph{(bert-sl256)} respectively. The intuition is that large scale pre-trained models are adept at language understanding and can be fine-tuned for new tasks with small number of samples. We finetune BERT for both datasets Yahoo! ans and Infobot respectively. It is non-trivial to include external information in BERT and we leave this for future work. 

\begin{table}[t]
\begin{tabular}{|l|l|l|l||l|l|l|}\hline
 & \multicolumn{3}{|c|}{Yahoo! Ans}  & \multicolumn{3}{|c|}{Infobot}  \\ \hline
Model & P$@$1 & R$@$3 & MRR  & P$@$1 & R$@$3 & MRR   \\ \hline
$\tau$-att &0.393	& 0.644 & 0.598 &0.673  &0.868  & 0.802\\ \hline
$\lambda$-sim  &0.3743	& 0.633 &0.578  &0.551 &0.817  &0.7207 \\\hline
bert-sl256 & 0.406 & 0.657 & 0.615	& 0.581 &  0.803 & 0.744 \\ \hline
bert-sl128 &0.363 & 0.604 &	0.589 & 0.557 &  0.799 & 0.731 \\ \hline
%ques-att \cite{} & &	&  &  & &\\ \hline
%lstm \cite{} & &	&  &  & &\\ \hline
%cnn+lstm \cite{} & &	&  &  & &\\ \hline
att & 0.377 & 0.645 &  0.589 & 0.567 & 0.821 & 0.739 \\ \hline
qasim &0.318 &0.608	& 0.546 & 0.551 &0.817 &0.720\\ \hline
random & 0.21 &- &-  &  0.239 & -& -\\ \hline
\end{tabular}
\caption{Evaluation with WHO external data}
\label{table:who_eval}
\end{table}
\subsection{Evaluation Metrics}
\label{subsec:eval_metrics}
We evaluate the performance of our model using three popular ranking metrics, mainly Precision (P$@$1), Mean Reciprocal Rank (MRR), and Recall (R$@$3). Each metric is described below:
\begin{itemize}
    \item \textbf{Precision (P$@$k):} Precision at position $k$ evaluates the fraction of relevant answers retrieved until position k. For, both datasets Yahoo! ans and Infobot \cite{emnlp_data2020}, we evaluate whether the top answer i.e. $(k=1)$ in the ranked list is indeed correct. It is defined as follows: 
        \begin{equation}
            Prec@k = \frac{1}{|Q|}\sum_{i=1}^{|Q|} \frac{\sum_{j=1}^{k} \mathbb{I}\{rel_{ij}=1\}}{k}
        \end{equation}
    where $\mathbb{I}\{rel_{ij}=1\}$ indicates whether the answer at position $j$ is relevant to the $i^{th}$ question. 
    \item \textbf{Recall (R$@$k):} Recall at position $k$ evaluates the fraction of relevant answers retrieved from all the answers marked relevant for a question. We report recall averaged for all the queries in test set. For recall, we take a cutoff as $(k=3)$, which evaluates whether the model is able to retrieve the correct answers in top 3 positions. It is defined as follows:
        \begin{equation}
            Recall@k = \frac{1}{|Q|} \sum_{i=1}^{|Q|} \frac{\sum_{j=1}^{k} \mathbb{I}\{rel_{ij}=1\}}{|rel_i|}
        \end{equation}
    where $|rel_i|$ is the number of relevant answers for the $i$th question. 
    \item \textbf{MRR (MRR):} evaluates the average of the reciprocal ranks corresponding to the most relevant answer for the questions in test set, which is given by:
        \begin{equation}
            MRR = \frac{1}{|Q|} \sum_{i=1}^{|Q|} \frac{1}{rank_i}
        \end{equation}
    where $|Q|$ indicates the number of queries in the test set and $rank_i$ is the rank of the \emph{first} relevant answer for the $i^{th}$ query.
\end{itemize}

\subsection{Parameter Settings}
Both primary datasets, Yahoo! ans and Infobot, were divied into three parts: train ($\sim$60\%), validation and test (20\%) respectively. 
The baseline models $\lambda$-sim and $att$ are initialized with glove embeddings \footnote{https://nlp.stanford.edu/projects/glove/} of 100 dimensions. We performed a parameter sweep over $\lambda$ and $\tau$ for $\lambda$-sim and $\tau$-att models with step size of 0.1 between $\{0, 1.0\}$ respectively. We used base uncased model for $bert$ implementation. We fine-tuned the model between 1-10 epochs and found that 3 epochs gave the best result on validation set. We used LSTM with 64 hidden units to represent question, answer and all the information in external datasets. We experimented with higher embedding size and hidden units, but the performance degraded significantly as the model tends to overfit on training data. Lastly we used batch size of 64 and trained the model for 30 epochs with early stopping. 

\begin{table}[t]
\begin{tabular}{|l|l|l|l||l|l|l|}\hline
 & \multicolumn{3}{|c|}{Yahoo! Ans}  & \multicolumn{3}{|c|}{Infobot}  \\ \hline
Model & P$@$1 & R$@$3 & MRR  & P$@$1 & R$@$3 & MRR   \\ \hline
$\tau$-att 	&0.532	&  0.778 & 0.715 & 0.606& 0.842 &0.766 \\ \hline
$\lambda$-sim  &0.326	& 0.616 &  0.555 & 0.556&  0.813& 0.722\\\hline
bert-sl256 & 0.406 & 0.657 & 0.615	& 0.581 &  0.803 & 0.744 \\ \hline
bert-sl128 &0.363 & 0.604 &	0.589 & 0.557 &  0.799 & 0.731 \\ \hline
att & 0.291 & 0.495 & 0.494 & 0.601 & 0.833 & 0.762 \\ \hline
%ques-att \cite{} & &	&  &  & &\\ \hline
%lstm \cite{} & &	&  &  & &\\ \hline
%cnn+lstm \cite{} & &	&  &  & &\\ \hline
qasim & 0.318&0.608	& 0.546 & 0.551 &0.817 &0.720\\ \hline
random & 0.21 &- &-  &  0.239 & -& -\\ \hline
\end{tabular}
\caption{Evaluation with TREC-COVID external data}
\label{table:trec_eval}
\end{table}
\section{Results} \label{sec:results}
In this work, our focus is to evaluate the utility of external information in improving answer ranking for cQ\&A task. Thus, we performed experiments to answer three main research questions listed below. \\
\textbf{RQ1:} Does external information improve answer ranking? \\ 
\textbf{RQ2:} How does temperature ($\tau$) compare with $\lambda$ parameter?  \\
\textbf{RQ3:} What kind of queries/questions does the model attend to when ranking relevant/non-relevant answers?  \\
\paragraph{\textbf{RQ1:} Does external information improve answer ranking?}
We evaluated different models for ranking answers in Yahoo! ans and Infobot dataset in presence of TREC and WHO datasets respectively. We found that temperature regulated attention models that incorporate external sources indeed outperform the baselines as shown in Table \ref{table:trec_eval} and Table \ref{table:who_eval} respectively. 
\begin{table}[t]
\begin{tabular}{|l|l|l|l|}\hline
Category & $\tau$-att & $\lambda$-sim &   qasim \\ \hline
Entertainment (47) &  0.829 & 0.702 & 0.59\\
Health (62) & 0.693& 0.69 & 0.645\\
Politics	(143) & 0.727& 0.629  &	0.587 \\
Society 	(38) & 0.578	 &  0.473 & 0.42 \\
Family (20) &0.85 & 0.750 & 	0.65 \\ \hline
\end{tabular}
\caption{Recall$@$3 of models across categories}
\label{table:cat_recall}
\end{table}
(\emph{$\tau$-att}) model beats \emph{bert} models by $\sim$30\% in precision, $\sim$18\% in recall and $\sim$16\% in MRR respectively on TREC data. However, (\emph{$\tau$-att}) does only marginally better than \emph{att} model in precision and MRR on Infobot data. We suspect that is due to the large set of query-document pairs in TREC-COVID data compared to fewer number of question-answer pairs in Infobot dataset. Our results also clearly suggest that embedding based matching of question-answer pair (\emph{qasim}) would not yield a good ranker, though it is better than choosing an answer at random (\emph{random}). When WHO is used as an external dataset, we find that (\emph{$\tau$-att}) model is slightly worse than \emph{bert}. This suggests that not all sources would equally benefit cQ\&A task. Since attention is dependent on the input query and key embedding lengths, it would be interesting to scale the computation in our model to incorporate several open external datasets to overcome this limitation in the future. 

Yahoo! ans questions are also assigned categories by users. Category based breakdown of performance on test set is given in Table \ref{table:cat_prec} and Table \ref{table:cat_recall} respectively, where categories with largest number of questions in test set are listed. In all the categories, our model outperforms best $\lambda$-sim and \emph{qasim} model respectively. The largest improvement happens for questions in Family  category where our model achieves an improvement of 71\% over the $\lambda$-sim model. It seems that ranking answers for questions from society and politics are harder than other categories. All the models, however, are able to rank the top answer in first three positions effectively as Recall$@$3 is high for all the categories. 
\begin{table}[t]
\begin{tabular}{|l|l|l|l|}\hline
Category & $\tau$-att & $\lambda$-sim &   qasim \\ \hline
Entertainment  (47) & 0.446& 0.382 & 0.297\\ 
Health (62) &0.483  &  0.419 &  0.354\\ 
Politics 	(143) & 0.45 & 0.300 &	0.272 \\
Society 	(38) & 	0.28 & 0.157 & 0.236 \\
Family  (20) & 0.6 & 0.350 & 0.40 \\  \hline
\end{tabular}
\caption{Precision$@$1 of models across categories}
\label{table:cat_prec}
\end{table}

\begin{table}[t]
\begin{tabular}{|l|l|l|l||l|l|l|}\hline
& \multicolumn{6}{c|}{\textbf{Temperature ($\tau$) > 1.0}} \\ \hline
  & 10 & 100 & 1000 & 10 & 100 & 1000 \\ \hline
\textbf{Src+ Ext} & \multicolumn{3}{|c||}{Prec$@$1} &  \multicolumn{3}{|c|}{Recall$@$3} \\ \hline
 Yahoo! + TREC & 0.46 &0.38 & 0.38& 0.73&0.644 & 0.64 \\ 
 Yahoo! + WHO& 0.37& 0.38& 0.36& 0.64& 0.65&0.64 \\
 Infobot + TREC & 0.44 & 0.59& 0.39& 0.72& 0.81& 0.75 \\
 Infobot + WHO & 0.65 & 0.41& 0.44 & 0.85 & 0.76 & 0.79\\ \hline
\end{tabular}
\caption{Variation in P$@$1 and R$@$3 across different temperature values.}
\label{table:temp_var}
\end{table}
 \begin{figure*}[t]
    \centering
    \begin{subfigure}[b]{0.49\columnwidth}
        \includegraphics[width=\textwidth]{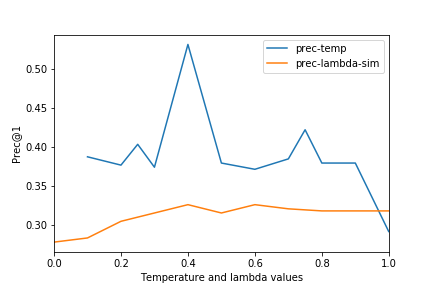}
    \caption{Yahoo!+TREC}
    \label{fig:yahoo_trec_temp}
    \end{subfigure}
    \begin{subfigure}[b]{0.50\columnwidth}
        \includegraphics[width=\textwidth]{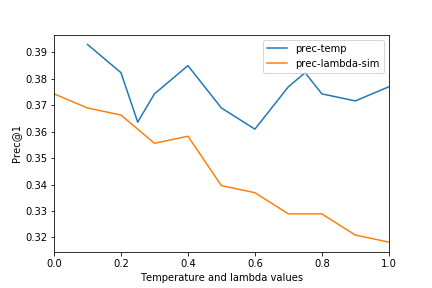}
        \caption{Yahoo!+WHO}
        \label{fig:yahoo_who_temp}
    \end{subfigure}
    \begin{subfigure}[b]{0.49\columnwidth}
        \includegraphics[width=\textwidth]{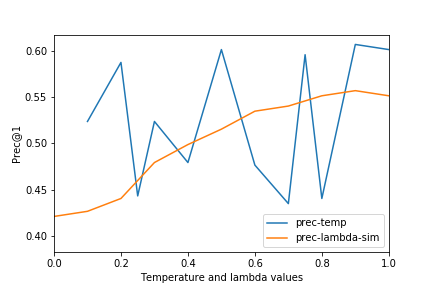}
        \caption{Infobot + TREC}
        \label{fig:emnlp_trec_temp}
    \end{subfigure}
     \begin{subfigure}[b]{0.49\columnwidth}
        \includegraphics[width=\textwidth]{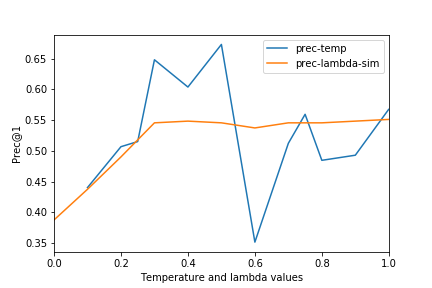}
         \caption{Infobot + WHO}
        \label{fig:emnlp_who_temp}
    \end{subfigure}
    \caption{Temperature and $\lambda$ variation impact on Prec$@$1}
    \label{fig:temp_var_img}
\end{figure*}

\paragraph{\textbf{RQ2:} How does temperature ($\tau$) compare with $\lambda$ parameter?}
We argued that linearly combining similarities between question-answer in primary dataset and between question-external source may not be sufficient to boost performance. We observe that in our results too i.e. $\lambda$-sim models do not perform better than (\emph{$\tau$-att}) models. This clearly indicates that more sophisticated models can learn to combine this information directly from training data. However, our experiments indicate that optimal value of (\emph{$\tau$}) varies across primary datasets and external sources of information. For instance, (\emph{$\tau$-att}) model performed best when \emph{$\tau=0.4$} and \emph{$\tau=0.9$} for Yahoo! ans and Infobot dataset respectively when TREC was used as external source. It performed best when \emph{$\tau=0.1$} and \emph{$\tau=0.5$} for Yahoo! ans and Infobot dataset respectively when WHO was used as external source. We also tried to vary \emph{$\tau$} beyond 1.0 to determine whether it yielded a trend as shown in Table \ref{table:temp_var}. Higher values of temperature seem to degrade model performance. We found that optimal temperature range is between $[0.1-1]$. Existing research in model distillation \cite{hinton2015distilling} has also empirically found that lower values of temperature yield better performance. 

We also compared model performance in terms of precision when $\lambda$ and $\tau$ are varied for $\lambda$-sim models and temperature based models respectively as shown in Figure \ref{fig:temp_var_img}. Temperature based models peak at one value but do not have a clear trend indicating that one needs to explore different $\tau$ values at the time of training for better performance. On the other hand, we observe that adding external information also helps the $\lambda$-sim models until a certain threshold. Overall, both sets of models show that free-text external information can be incorporated to improve answer ranking performance. 

\begin{figure}[t]
  \includegraphics[width=0.50\textwidth]{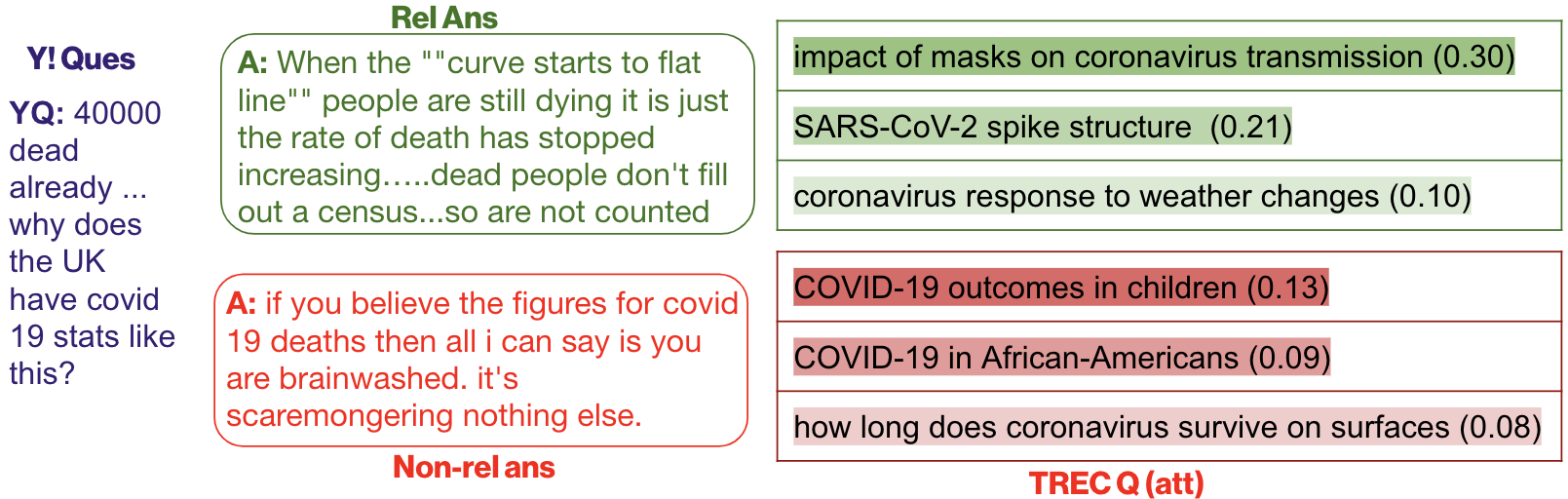}
  \caption{Y! ques, its rel and non-rel ans and questions with $\tau$-att model's attention values for TREC queries.}
  \label{fig:yahoo_att_example}
\end{figure}

\begin{figure}[t]
  \includegraphics[width=0.50\textwidth]{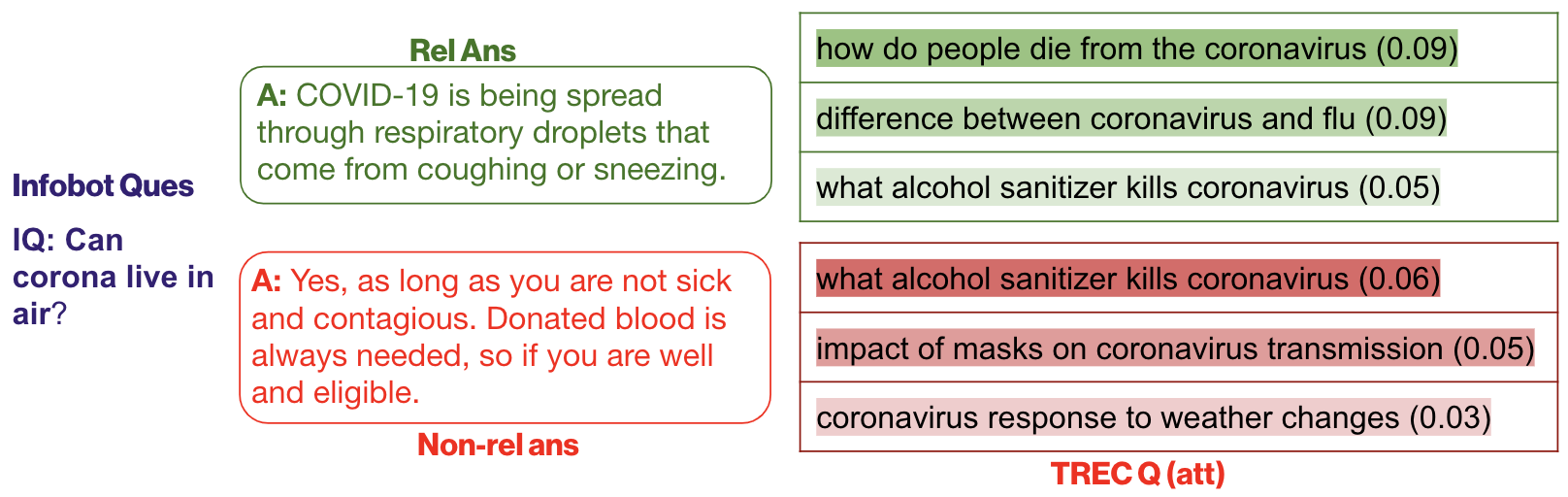}
  \caption{Infobot ques, its rel and non-rel ans and questions with $\tau$-att model's attention values for TREC queries.}
  \label{fig:infobot_att_example}
\end{figure}

\paragraph{\textbf{RQ3:} What kind of queries/questions does the model attend to when ranking relevant/non-relevant answers?}
Attention based models have a very unique feature: they can aid explaining the internal workings of neural network models. We inspect what kind of queries/questions in external datasets does our model pay attention to while ranking relevant or non-relevant answers. Figure \ref{fig:yahoo_att_example} shows one such example of Yahoo! question and incorporation of TREC data. At the time of scoring relevant answer, the model gives higher weight to some queries compared to others. In the example, for instance, it assigns more weight to queries associated with masks or COVID virus response to weather changes. We observe higher attention weights for questions when relevant answers are ranked than when non-relevant answers are scored. An example question, a relevant and non-relevant answer along with model attention weights on TREC queries are shown from the Infobot data in Figure \ref{fig:infobot_att_example} respectively. It shows a similar trend where attention weights are high for external queries that are closely associated with the question answer text. 

Overall, our experiments show that curated external information is useful for improving community question answering task. Our experiments also indicate that this external knowledge need not always be structured text. However, it is worth noting that curated and reliable external sources may not always be available for all domains. We addressed a very niche task in this work, and further research is required to extend it to incorporate multiple external sources.
We posit that with scalable attention mechanisms, this work can be easily made tractable for large external sources containing thousands or millions of entries in the future. 
%It is interesting to note that $\lambda$-sim model achieves highest increment in P$@$1 over question-answer matching model (\emph{yqssim}) for questions in \emph{Entertainment} category which further confirms the utility of adding external information to our matching models. 

%The precision and recall performance of each model is given in Table \ref{table:prec_recall}. We list performance of our model (\emph{$\tau$-att}) when trained with certain values of attention. As mentioned before $t=\{0.25, 0.5, 0.75,1.0,2.0, 10.0\}$ for this experiment. We also vary $\lambda$ for $\lambda$-sim models in range of $\{0.25, 0.5, 0.75, 1.0\}$.  

%For $\lambda$-sim models, we find that 0.5 yields better results which indicates that \emph{both} Yahoo! question and trec content based similarity is useful for finding the top answer. Our model with temperature $t=0.25$ performs the best among all values of temperature. While conceptually our model attempts to find the most effective combination of Yahoo! question and trec information for ranking answers, it beats the crude linear combination of both by 26\% in precision and 17\% in recall. It was interesting to note that higher values of temperature did not improve model's performance. 

% TREC
%$\tau$-att -- yahoo (0.4), emnlp (0.9)
% lambda-sim -- yahoo (0.6), emnlp (0.9)

% WHO
%$\tau$-att -- yahoo (0.1), 0.5
% lambda-sim -- yahoo (0.0), 1.0

\section{Conclusion}
Question answering platforms provide users with effective and easy access to information. These platforms also provide content on rapidly evolving \emph{sensitive} topics such as disease outbreaks (such as COVID-19) where it is also useful to use external \emph{vetted} information for ranking answers. Existing work only exploits knowledge bases which have some limitations that makes it difficult to use them for community Q\&A for rapidly evolving topics such as wild-fires or earthquakes. In this work, we tried to evaluate the effectiveness of external (free text or semi-structured) information in improving answer ranking models. We argue that simple question-answer text matching may be insufficient and in presence of external knowledge, but temperature regulated attention models can distill information better which in turn yields higher performance. Our proposed model with temperature regulated attention, when evaluated on two public datasets showed significant improvements by augmenting information from two \emph{external} curated sources of information. In future, we aim to expand these experiments to other categories such as disaster relief and scale the attention mechanism to include multiple external sources in one model.

\bibliographystyle{abbrv}
\bibliography{www2021}

\end{document}